\pdfoutput=1

\documentclass[11pt]{article}

\usepackage[final]{latex/acl}

\usepackage{times}
\usepackage{latexsym}

\usepackage[T1]{fontenc}

\usepackage[utf8]{inputenc}

\usepackage{microtype}
\usepackage{phaistos}
\usepackage{soul}
\usepackage{inconsolata}

\usepackage{graphicx}
\usepackage{array}

\usepackage{amsmath}

\usepackage{multirow}
\usepackage{multicol}

\newcommand{\re}[1]{{\color{black}#1}}
\title{\PHtunny \, Fishing for Answers: Exploring One-shot vs. Iterative Retrieval Strategies for Retrieval Augmented Generation}

\author{Huifeng Lin, Gang Su, Rui Zhao \\
  SenseTime Research \\
  \texttt{zhaorui@sensetime.com} \\
  \And
  Jintao Liang \\
  BUPT, Beijing \\
  \texttt{ljt2016@bupt.edu.cn} \\
  \AND
  You Wu \\
  HKUST, Hong Kong \\
  \texttt{wuyouscut@gmail.com}\\
  \And
  Ziyue Li \\
  Technical University of Munich, Germany \\
  \texttt{ziyue.li@tum.de}}

\begin{document}
\maketitle
\begin{abstract}
Retrieval-Augmented Generation (RAG) based on Large Language Models (LLMs) is a powerful solution to understand and query the industry's closed-source documents.
However, basic RAG often struggles with complex QA tasks in legal and regulatory domains, particularly when dealing with numerous government documents.
The top-$k$ strategy frequently misses golden chunks, leading to incomplete or inaccurate answers.
To address these retrieval bottlenecks, we explore two strategies to improve evidence coverage and answer quality. The first is a One-SHOT retrieval method that adaptively selects chunks based on a token budget, allowing as much relevant content as possible to be included within the model's context window. Additionally, we design modules to further filter and refine the chunks. The second is an iterative retrieval strategy built on a Reasoning Agentic RAG framework, where a reasoning LLM dynamically issues search queries, evaluates retrieved results, and progressively refines the context over multiple turns. We identify query drift and retrieval laziness issues and further design two modules to tackle them. 
Through extensive experiments on a dataset of government documents, we aim to offer practical insights and guidance for real-world applications in legal and regulatory domains.
The code and dataset will be released upon acceptance.

\end{abstract}

\section{Introduction}

\re{Retrieval-Augmented Generation (RAG)~\cite{chen2024benchmarking,lewis2020retrieval,gao2023retrieval} based on Large Language Models (LLMs) has been intensively deployed to conduct question-answering (QA) towards closed-source or internal documents.} Although Large Language Models (LLMs)~\cite{singh2023exploring,zhao2023survey,zhu2024large} have demonstrated remarkable capabilities in natural language understanding and generation, LLMs rely on static training data, making them prone to hallucinations and limiting their ability to provide accurate, up-to-date information in knowledge-intensive tasks~\cite{rawte2023troubling,zhang2023siren,huang2025survey}.
By integrating relevant information from external knowledge bases or search engines, RAG enhances factual accuracy and broadens the model's domain coverage \cite{zhao2024retrieval,li2024enhancing}. 
\begin{figure}[t]
  \includegraphics[width=\columnwidth]{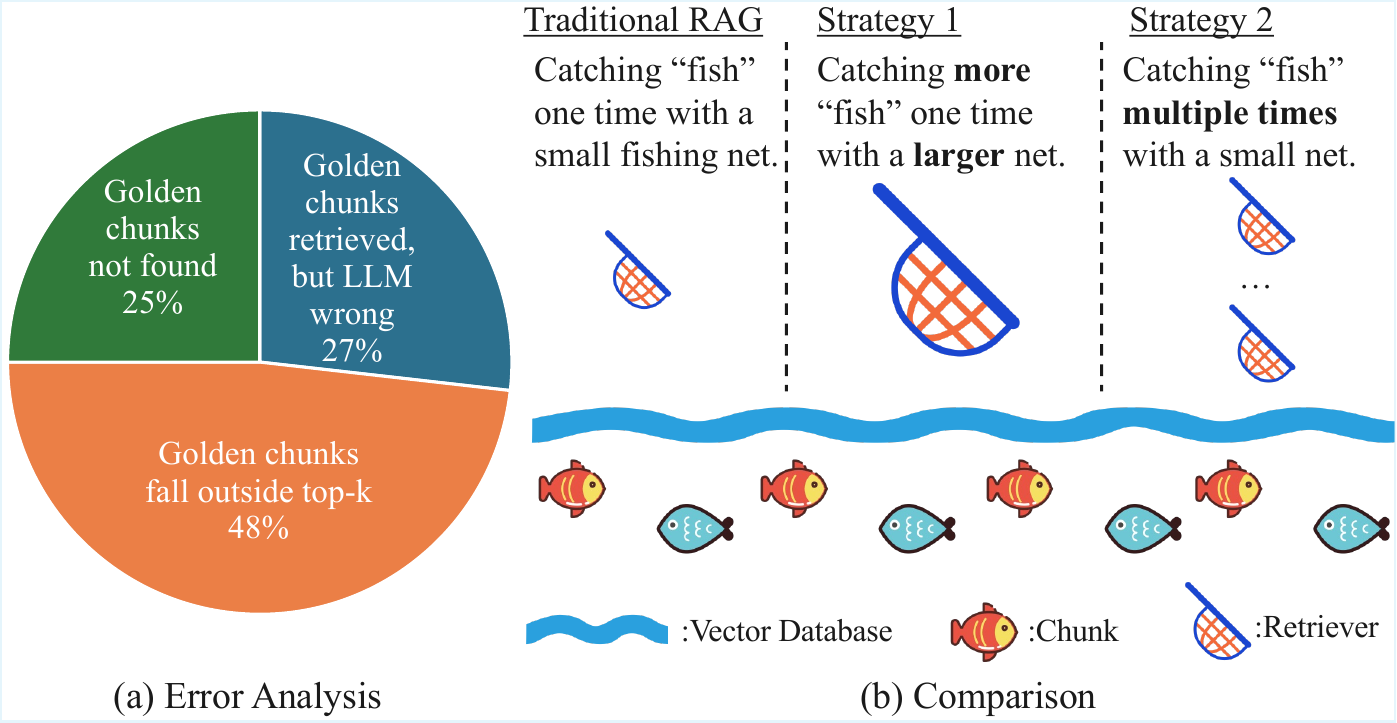}
  \vspace{-20pt}
  \caption{(a) Error analysis: 48\% of traditional RAG failures is due to not finding the golden chunk in the top-$k$; (b) Comparison of three retrieval methods: Traditional RAG is casting a small fishing net once; We propose either to use a ``larger'' net or a same net but ``multi-times''.}
  \vspace{-15pt}
  \label{fig:fishing}
\end{figure}

\re{Traditional RAG methods typically follow a one-round retrieval paradigm, where the user’s query is first embedded into a vector representation and matched against a vector-based index of document chunks stored in a knowledge base or internal repository. The retriever then selects the top-ranked chunks based on vector similarity, and these retrieved chunks, together with the original query, are provided to a generative LLM, which synthesizes the final answer using both the retrieved evidence and the query context.}


\re{Despite the effectiveness of basic RAG methods, they often struggle when applied to real-world scenarios involving complex and heterogeneous government documents. In our analysis of QA tasks on such documents (Fig. \ref{fig:fishing}(a)), we find that nearly half of the incorrect answers result from the failure to retrieve the golden chunks using the traditional top-$k$ selection strategy. This retrieval failure prevents the generative model from accessing crucial evidence, ultimately degrading the quality of the final answer.} 

\re{Since the one-round retrieval may miss the golden chunks, then the intuitive practical solution is: (1) either we increase the number of chunks retrieved, or (2) keep the same number of chunks and just retrieve multiple times. Take the analogy of fishing (Fig. \ref{fig:fishing}(b)), the traditional RAG's top-k retrieval can be seen as casting a small fishing net once (e.g., top-5 usually), hoping to capture all relevant evidence in a single pass. However, this “single-shot trawling” approach may miss the golden chunks altogether, especially in complex or diverse knowledge domains. The first solution is basically to dase a larger net, but it may catch too much irrelevant information (“noise” or “junk fish”); The second is to adopt a “multi-cast” fishing strategy, where smaller, more targeted nets are cast repeatedly, progressively refining the retrieval toward high-value chunks.} 


Motivated by the two retrieval ideas introduced above: casting a larger net and casting multiple times, we implement two corresponding strategies. The first is a One-SHOT retrieval strategy that removes the fixed top-$k$ constraint and instead selects as many relevant chunks as fit within a predefined token budget, ranking them by relevance-per-token to maximize evidence density. To further enhance precision, we incorporate rule-based modules, such as \re{\textit{chunk filter} to filter out irrelevant chunks based on chunk meta information (e.g., year, location in our case).} The second is an iterative retrieval strategy based on an Agentic RAG framework, where a reasoning-capable LLM manages retrieval over multiple turns by issuing intermediate queries, assessing retrieved results, and progressively refining the context. These strategies offer practical solutions to improve evidence coverage and robustness in complex QA scenarios.


\re{The complex QA usually lies in two abilities of RAG system, which we refer to as the reasoning ability: (1) the ability to decompose the complex question (generation reasoning), and (2) the ability to retrieve multiple cues (retrieval reasoning). To systematically evaluate the performance of traditional RAG systems and our two proposed strategies on government documents, we design a benchmark that categorizes questions into four levels based on whether generation reasoning and retrieval reasoning are required.}

\re{This paper offers valuable insights (underlined) through extensive experiments on real-world government documents, which we believe are highly beneficial for applications in legal and regulatory domains. The resulting RAG system has been continuously deployed in practice and serving various clients for over two years.}

\section{Related Works}
\label{related work}

\subsection{Basic RAG}
\textit{Retrieval-Augmented Generation} (RAG) was introduced to overcome the static knowledge limitations of LLMs by integrating external retrieval mechanisms during inference~\cite{chen2024benchmarking,gao2023retrieval}. Naive RAG methods represent the earliest implementations, typically using sparse retrieval techniques like BM25~\cite{robertson2009probabilistic} to fetch documents based on keyword overlap~\cite{ma2023query}. While efficient for simple factoid queries, these approaches offered limited semantic understanding, thus often retrieving noisy or redundant content and failing to reason across multiple sources.

The emergence of Advanced RAG and Modular RAG was aimed at addressing key limitations of the Naive RAG, particularly in terms of retrieval precision, information integration, and system flexibility~\cite{gao2023retrieval}. 
Advanced RAG improves retrieval quality through techniques such as dense semantic matching, re-ranking, and multi-hop querying, while also introducing refined indexing strategies like fine-grained chunking and metadata-aware retrieval. 
Modular RAG rethinks the Naive RAG by breaking down the end-to-end process of indexing, retrieval, and generation into discrete, configurable modules.
This design allows for greater architectural flexibility and enables system developers to incorporate diverse techniques into specific stages, such as enhancing retrieval with fine-tuned search modules \cite{lin2023ra}.
In response to specific task demands, various restructured and iterative module designs have also emerged. As a result, modular RAG has increasingly become a dominant paradigm in the field, supporting both serialized pipeline execution and end-to-end learning across modular components.

Despite these advances, many RAG systems remain constrained by static control logic, making them ill-suited for complex QA tasks where key evidence may be scattered or initially missed. Recent work on Agentic RAG introduces reasoning and tool use into retrieval, enabling more adaptive behavior \cite{ravuru2024agentic,li2025searcho1agenticsearchenhancedlarge,wu2025agenticreasoningreasoningllms}.
Motivated by this, we explore two complementary strategies: a One-SHOT retrieval method with token-aware evidence selection, and an agent-driven iterative framework. These approaches aim to improve retrieval robustness and adaptivity in real-world QA scenarios.

\subsection{Agentic RAG}
The year 2025 is marked as the year of agentic AI, with applications emerging such as agentic LLMs and so on \cite{ruan2023tptu,kong2024tptu,zhang2024controlling}. Recent progress in RAG has moved beyond static, rule-based pipelines toward more dynamic, decision-driven systems broadly known as \textit{Agentic RAG}. These systems embed retrieval decisions into the model’s reasoning flow, enabling LLMs to actively determine when and how to interact with external tools during generation. 

A growing body of work has demonstrated the effectiveness of prompting large models to autonomously invoke tools, reformulate queries, or break down complex questions. For instance, methods like ReAct~\cite{yao2023reactsynergizingreasoningacting}, Self-Ask~\cite{press2023measuringnarrowingcompositionalitygap}, and Search-o1~\cite{li2025searcho1agenticsearchenhancedlarge} allow models to interleave generation with retrieval by identifying knowledge gaps and issuing targeted queries. Built-in function calling APIs~\cite{openai_function_calling} further support structured tool use in models like GPT and Gemini.

Beyond prompting, other approaches train models to improve their retrieval capabilities through learning signals. DeepRetrieval~\cite{jiang2025deepretrievalhackingrealsearch}, Search-R1~\cite{jin2025searchr1trainingllmsreason}, and ReZero~\cite{dao2025rezeroenhancingllmsearch} integrate retrieval into the learning loop, optimizing search behaviors with reinforcement signals. More advanced systems like DeepResearcher~\cite{zheng2025deepresearcherscalingdeepresearch} extend this idea to open-domain web environments, encouraging models to reason, search, and synthesize in a tightly coupled loop.

\section{Methodology}
This section will  give the details on how we achieve the two strategies.

\subsection{One-SHOT Strategy: \re{``Casting A Bigger Net!''}}
\label{sec:one_shot_strategy}
Our One-SHOT (all capital to infer the bigger net) retrieval strategy implements a token-constrained RAG framework designed to maximize recall within a fixed context window. Rather than selecting a fixed number of top-$k$ (e.g., top-5) chunks, the system dynamically selects \re{as many as possible} relevant chunks that can fit within a given token budget, enhancing evidence coverage in a single retrieval, \re{denoted as Token-Constrained Top-$K_{max}$}. A few components are further designed:

\begin{figure}[t]
  \includegraphics[width=\columnwidth]{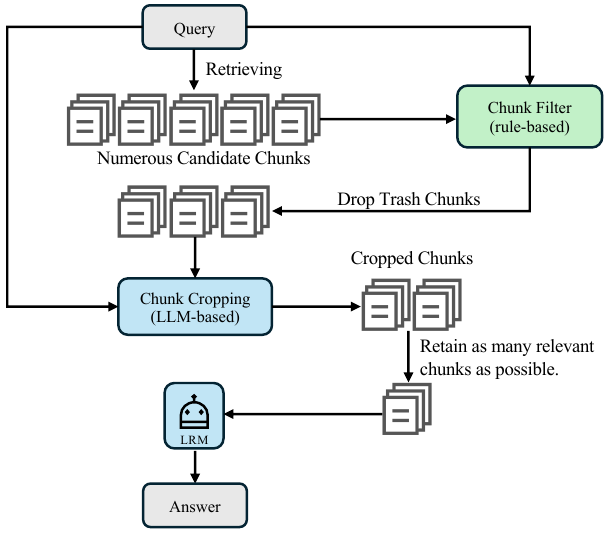}
  \vspace{-20pt}
  \caption{One-SHOT Strategy. This category includes methods such as Token-Constrained Top-K (retain as many relevant chunks as possible), along with techniques like chunk filter and chunk cropping. 
  }
  \vspace{-10pt}
  \label{fig:one-shot_strategy}
\end{figure}

\subsubsection{Token-Constrained Top-$K_{max}$}

To mitigate the issue of golden chunks being excluded under a fixed Top-$k$ setting, we propose a Token-Constrained Top-$K_{max}$ strategy that aims to \textbf{maximize recall within a predefined token budget} (e.g., 32,000 tokens). Each candidate chunk is associated with a relevance score $r_i$ (e.g., embedding similarity) and a token count $t_i$, and the goal is to select a subset of chunks that together fit within the context window while maximizing the total relevance. Let $x_i \in \{0,1\}$ be a binary variable indicating whether chunk $i$ is selected, and $T_{\text{max}}$ be the token limit. The selection problem can be formalized as:
\begin{equation}
    \begin{aligned}
       & \max \sum_{i=1}^n r_i \cdot x_i \\
       & \text{s.t.} \quad \sum_{i=1}^n t_i \cdot x_i \leq T_{\text{max}}, \quad x_i \in \{0,1\}
    \end{aligned}
\end{equation}
This formulation effectively prioritizes chunks with high relevance-per-token ratios, enabling dense and informative context construction within the model’s input limits.









\subsubsection{Key Optimization Components}

\re{As illustrated in Fig. \ref{fig:fishing}(b), such a greedy retrieval inevitably introduce some unrelated chunks.} To further refine the evidence retrieved by the Token-Constrained Top-$K_{max}$ strategy, we design two complementary optimization components: \textit{chunks filter} and \textit{chunks summary}. 

The \textbf{chunk filter} component \re{(rule-based)} selectively removes irrelevant chunks and augments missing but potentially useful ones by analyzing \re{chunk meta information} such as temporal expressions, locations, and named entities. Specifically, the drop operation filters out chunks that lack alignment with query entities, while the add operation supplements the retrieved set with additional chunks containing matching or complementary elements. This rule-based filtering improves both precision and recall of the retrieval results.

The \textbf{chunk cropping} component (LLM-based) \re{further crops the remaining related chunks. It} leverages LLM reasoning to assess and refine the initial evidence holistically. \re{Given a specific related chunk that survives the filtering process, the LLM evaluates the contents inside this chunk carefully with respect to the original query and crop out the information deemed irrelevant or redundant. Chunks are basically refined to be shorter after this process}, resulting in a more condensed and informative evidence set for answer generation. 

\re{In practice, both chunk filter and chunk cropping are conducted first before the token length is satisfied, such that the recall is being possibly maximized.}

\subsection{Iterative Retrieval Strategy: \re{``Casting A Small Net Multi-Times!''}}
\label{sec:iterative_strategy}
Our iterative retrieval strategy implements a reasoning agentic RAG \cite{liang2025reasoning}  that leverages the reasoning and function-calling capabilities of large reasoning language models (LRMs) to perform adaptive information retrieval.

\subsubsection{Architecture Overview}

\begin{figure}[t]
  \includegraphics[width=1.02\columnwidth]{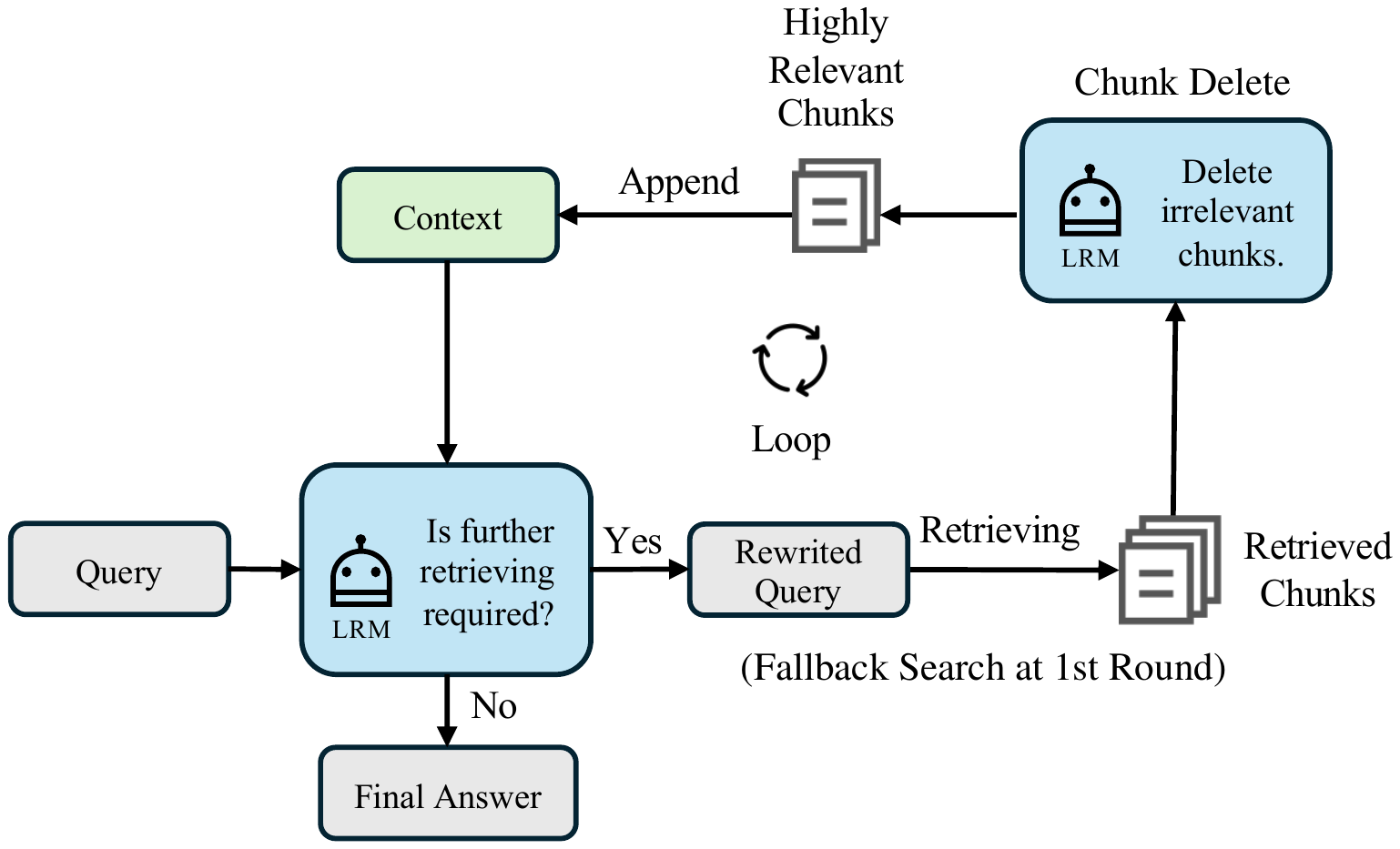}
  \vspace{-20pt}
  \caption{Iterative retrieval strategy, with fallback module to avoid query drift and chunk delete to avoid retrieval laziness.}
  \vspace{-10pt}
  \label{fig:iterative_retrieval_strategy}
\end{figure}

As shown in Fig. \ref{fig:iterative_retrieval_strategy}, the iterative retrieval system follows a multi-turn paradigm where the language reasoning model acts as an intelligent agent capable of making informed decisions about when and how to retrieve additional information. Unlike traditional RAG systems that rely on explicit pipeline components, our approach embeds the core functionalities implicitly within the model's reasoning process, managing context analysis, tool invocation, and information synthesis naturally through the model's inherent reasoning abilities.

\re{Similar to existing agentic RAG models, e.g., Search-O1 \cite{li2025search} and Search-R1 \cite{jin2025search}}, the system employs a structured reasoning process encapsulated within \texttt{<think>} and \texttt{</think>} tags, where the model evaluates information sufficiency, identifies knowledge gaps, and formulates search queries. When additional information is required, the model generates structured function calls using the \texttt{<tool\_call>} format. The system provides two primary tools: \texttt{chunk\_search} for information retrieval (\re{still use the small ``fishing net''}) and \texttt{chunk\_delete} for context refinement and information overload prevention.

\subsubsection{Multi-Turn Retrieval Process}

The iterative retrieval process operates within a configurable maximum turn limit (typically set to 5 turns) to balance thoroughness with efficiency. In each turn, the system performs the following steps:

1. \textbf{Reasoning Phase}: The model analyzes the current context, including the original query and all previously retrieved information, to determine if additional retrieval is necessary.

2. \textbf{Query Formulation}: If retrieval is deemed necessary, the model generates a focused search query designed to address specific information gaps identified during reasoning.

3. \textbf{Retrieval Execution}: The system executes the \texttt{chunk\_search} function against the vector database to identify relevant document chunks. The function accepts the rewritten query and returns the top-$k$ most relevant document chunks ($k=5$) with metadata and text content.

4. \textbf{Context Integration}: Retrieved chunks are integrated into the system context, providing the model with additional information for subsequent reasoning.

5. \textbf{Iteration Decision}: The model determines whether to continue searching or provide a final answer based on the accumulated information.

\subsubsection{Key Optimization Components}

Our complete iterative system incorporates several critical optimization mechanisms that address the inherent challenges of agentic retrieval:

\textbf{Fallback Search Mechanism}: To mitigate the risk of query drift and ensure comprehensive coverage, the system implements a strategic fallback mechanism during the initial retrieval phase. When the model performs its first \texttt{chunk\_search} operation, the system automatically executes an additional retrieval using the original user query in parallel to the model's reformulated query. This dual-retrieval approach is specifically applied only to the first turn to provide a strong foundation for subsequent iterations—ensuring the model starts with a good baseline of relevant information from the original query while still benefiting from its own query reformulation. 

\textbf{Adaptive Context Management}: The system incorporates the \texttt{chunk\_delete} function, which allows the model to remove irrelevant retrieved information from the working context by specifying the chunk IDs of unwanted chunks. This mechanism helps maintain focus on pertinent information and prevents the accumulation of noise that could degrade answer quality.

These optimization components are essential for addressing two key challenges: 
\begin{itemize}
    \item \re{\textbf{Query drift}}: where autonomous query reformulation by LRM drifts away from the original query and leads to less relevant results, as shown in Appx. \ref{appendix:query_drift}.
    \item \textbf{Retrieval laziness}: where in the iterative agentic retrieving, the model prematurely terminates search due to context overload: specifically, the heavier the context retrieved from previous round, the less likely the agent will initiate the next retriecal action due to ``cognitive burden'', as shown in Appx. \ref{appendix:retrieval_laziness}. Such a cognitive burden is also observed in function calling \cite{yang2025pencil}.
\end{itemize}

The effectiveness of these components is demonstrated in our experiments in Section~\ref{sec:iterative_experiments}.




\section{Experiment}
\subsection{Experiment Settings}
\begin{figure}[h]
  \includegraphics[width=\columnwidth]{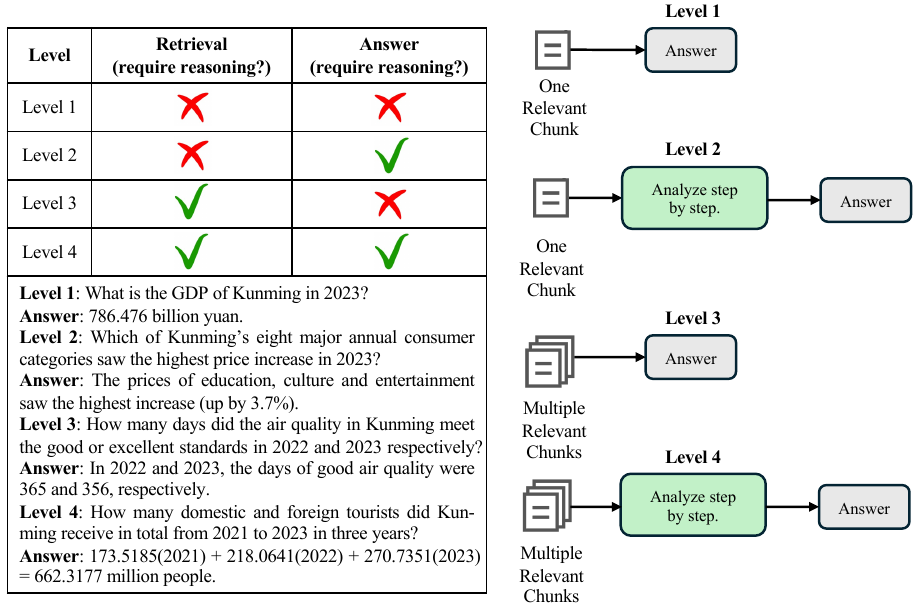}
  \vspace{-20pt}
  \caption{Categories of questions in our dataset. The table in the upper-left shows differences between questions of various levels. The lower-left provides examples of different-level questions and their reference answers. The diagrams on the right illustrate the solving processes for questions of different levels.}
  \label{fig:question_example}
\end{figure}

\textbf{Datasets}. Our dataset is composed of 1,000 carefully designed questions, all extracted from a corpus of over 40,000 real-world government documents. Specifically, the questions are categorized into four levels (ranging from Level 1 to Level 4) based on two key criteria: whether retrieval requires reasoning and whether answering requires reasoning. We ensure that all answers are retrievable from the original document corpus. Figure \ref{fig:question_example} illustrates the difference between each level.


\textbf{Basic RAG Setting}. Our baseline system employs traditional single-turn retrieval with a fixed Top-5 strategy. This basic RAG approach retrieves the top-5 most relevant document chunks based on vector similarity and provides them to the language model for answer generation without any iterative refinement or optimization mechanisms. All experiments are conducted using Qwen3-32B as the language model.

\textbf{Evaluation Setting}. We employ an LLM-as-a-judge evaluation framework to assess answer quality across multiple dimensions including factual accuracy, completeness, and relevance.
We adopt SenseChat-5 to perform answer evaluation.
The evaluation scores are normalized to a 0-100 scale for consistent comparison across different approaches. 

\begin{table}[t]
\centering
\resizebox{1.0\columnwidth}{!}{
\begin{tabular}{l|c|c|c|c|c}
\hline
\textbf{Method} & \textbf{L1} & \textbf{L2} & \textbf{L3} & \textbf{L4} & \textbf{Avg} \\
\hline
Basic RAG (Top-5)                & 89.5 & 88.5 & 76.0 & 70.0 & 81.0 \\
\hline
Token-Constrained Top-$K_{max}$  & 90.0 & 91.0 & 85.5 & 83.5 & 87.5 \\
\quad + Chunk Filter             & \textbf{92.0} & \textbf{93.0} & \textbf{90.0} & \textbf{89.0} &\textbf{91.0} \\
\quad + Chunk Cropping            & 91.0 & 92.5 & 86.5 & 84.0 & 88.5 \\
\quad + Chunk Filter \& Cropping  & 92.0 & 93.0 & 87.5 & 83.5 & 89.0 \\
\hline
\textbf{Improvement over Baseline} & \textbf{+2.5} & \textbf{+4.5} & \textbf{+14.0} & \textbf{+19.0} & \textbf{+10.0} \\
\hline
\end{tabular}
}
\vspace{-5pt}
\caption{Ablation study of One-SHOT retrieval components. Each row builds incrementally on the previous configuration.}
\label{tab:oneshot_results}
\vspace{-10pt}
\end{table}

\subsection{Results on One-SHOT Strategy}
\label{sec:oneshot_experiments}

\subsubsection{Ablation studies}

We first conduct ablation studies on the One-SHOT retrieval strategy to evaluate the impact of each optimization component, as shown in Table~\ref{tab:oneshot_results}. 
Starting from the baseline \textit{Basic RAG (Top-5)}, our \textit{Token-Constrained Top-$K_{max}$} strategy brings a notable improvement \ul{by relaxing the fixed top-$k$ constraint and retrieving more relevant content within a token budget}. This approach performs particularly well on L3 and L4 tasks, which demand deeper reasoning and broader evidence coverage. With the addition of \textit{Chunk Filter}, we achieve the highest overall performance (91.0\% average), amounting to a +10.0\% gain over the baseline. This confirms the effectiveness of combining token-based recall expansion with relevance-aware filtering.


Adding the \textit{Chunk Filter} before reaching the token budgets further boosts performance, achieving the best overall result (91.0\% average). The largest gains appear on L3 (+14.0\%) and L4 (+19.0\%), where \ul{filtering irrelevant chunks mitigates noise from large retrieval sets and allows the model to focus on high-value evidence.} 
The \textit{Chunk Cropping} further improves performance by removing noisy chunks through LLM-based semantic judgments. However, its effect is more moderate compared to the rule-based \textit{Chunk Filter}, possibly due to the \ul{trade-off between precision and recall introduced by aggressive semantic cropping}. 
Combining \textit{Chunk Filter} and \textit{Chunk Cropping} yields mixed results, suggesting that their benefits are not strictly additive and require careful integration.

Overall, these results demonstrate that optimizing evidence selection through both token budgeting and relevance refinement is crucial for enhancing RAG performance in complex QA scenarios.

\subsubsection{Chunk Retention Analysis}
To better understand how the One-SHOT pipeline refines evidence through multi-stage processing, we analyze the \textit{chunk retention ratio} at each stage: after the initial retrieval (Token-Constrained Top-$K_{max}$), after applying the rule-based \textit{Chunk Filter}, and finally after applying LLM-based \textit{Chunk Cropping}. 
The results in Table~\ref{tab:chunk_retention_updated} show that chunk retention varies across QA difficulty levels, with L2 consistently yielding the lowest retention. Both \textit{Chunk Filter} and \textit{Chunk Cropping} reduce chunk volume, with the latter being slightly more aggressive across all levels. Interestingly, L1 and L4 retain more chunks after refinement, suggesting that \ul{simple factoid queries (L1) and domain-specific regulatory queries (L4) allow clearer chunk-to-query alignment}.
When combined, the two methods further reduce context size while maintaining high relevance, confirming that \ul{structured and semantic filtering are complementary in eliminating irrelevant or low-utility content}, and that refinement sensitivity varies by question type.

\begin{table}[h]
\centering
\resizebox{1.0\columnwidth}{!}{
\begin{tabular}{l|cccc}
\hline
\textbf{Methods} & \textbf{L1} & \textbf{L2} & \textbf{L3} & \textbf{L4} \\
\hline
+ Chunk Filter                &85 & 76    &80  &82 \\
+ Chunk Cropping             & 83 & 73    &75  &79  \\
+ Chunk Filter \& Cropping   & 80 & 71    &72  &76   \\
\hline
\end{tabular}
}
\caption{Chunk retention ratio ($\%$) after applying chunk refinement modules under different QA difficulty levels. Values indicate the percentage of retrieved chunks retained after each processing step.}
\label{tab:chunk_retention_updated}
\end{table}

\begin{table}[h]
\centering
\resizebox{1.0\columnwidth}{!}
{
\begin{tabular}{c|c|cc}
\hline
\textbf{Models} &\textbf{Metrics} & \textbf{Basic RAG (Top-5)} &\textbf{One-SHOT Strategy}  \\
\hline

\multirow{2}{*}{SenseChat-5} & Scores  &77.5  &83.5 \\
                             & Mins    &30     &24 \\
\hline

\multirow{2}{*}{Qwen2.5-32B} & Scores  &\textbf{80.5}   &\textbf{91.5} \\
                             & Mins    &23     &21 \\
\hline

\multirow{2}{*}{Qwen3-32B no think} & Scores  &77.0   &91.0 \\
                             & Mins    &26     &24 \\
\hline

\multirow{2}{*}{Qwen3-32B think} & Scores  &75.0   &89.5 \\
                             & Mins    &51     &83 \\
\hline
\end{tabular}
}
\caption{Evaluation scores and time costs on different LLMs for One-SHOT strategy.}
\label{tab:different_LLMs}
\end{table}

\subsubsection{Applications on Different LLMs}
\re{As One-SHOT strategy basically increase the recall, so theoretically it won't behave too much differently on L1 to L4, instead, we would like to see how the different choice of LLMs will affect the overal result, thus,} Table~\ref{tab:different_LLMs} only presents the overall evaluation scores and time costs of applying the One-SHOT strategy across different LLMs. Compared to the baseline \textit{Basic RAG (Top-5)}, \ul{all models show substantial performance improvements when using ``a big net'', i.e., the token-constrained One-SHOT strategy}. For instance, \textit{Qwen2.5-32B} achieves the highest score of 91.5 (+11.0), while \textit{SenseChat-5} improves by 6.0 points to reach 83.5.
In terms of efficiency, the One-SHOT strategy generally reduces time costs for most models, as it avoids multi-turn reasoning and minimizes redundant computations. For example, the time for \textit{SenseChat-5} drops from 30 to 24 minutes. However, for \textit{Qwen3-32B} in the \textit{think} configuration, we observe a significant increase in latency (from 51 to 83 minutes). This overhead is attributed to the additional reasoning steps performed within the \texttt{<think>} blocks during relevance evaluation.
These results demonstrate the effectiveness and generality of the One-SHOT strategy across LLMs with different sizes and reasoning capabilities. They also highlight the trade-off between reasoning complexity and latency, suggesting that careful system design is needed to balance performance gains and computational efficiency.


\subsection{Results on Iterative Retrieval Strategy}
\label{sec:iterative_experiments}

\subsubsection{Ablation studies}

We evaluated our iterative retrieval approach through a series of ablation studies, progressively building from a basic agentic system to our complete framework described in Section~\ref{sec:iterative_strategy}. Our experiments reveal several key insights about the challenges and opportunities in multi-turn retrieval systems.
As shown in Table \ref{tab:iterative_results}, the \textit{Basic Agentic (Multi-turn)} can autonomously decide when and what to search across multiple turns, \ul{but it shows mixed results: while complex retrieval tasks (L3 and L4) benefit from the iterative approach, simpler questions (L1 and L2) experience slight performance degradation}. This counterintuitive result reveals a critical challenge in agentic retrieval: \textbf{query drift}, where the model's autonomous query reformulation may deviate from the user's original intent, particularly for straightforward questions where the initial query is already well-formed. 
For example, when given the original query "I left my previous company 3 years ago, can I still recover my housing fund contributions?", the model might reformulate it as "housing fund withdrawal time limit regulations" and then further drift to "housing fund withdrawal conditions for resigned employees". This reformulation fundamentally changes the intent from recovering/reclaiming contributions to asking about withdrawal procedures, leading to irrelevant retrieval results.
\begin{table}[t]
\centering
\resizebox{1.0\columnwidth}{!}{
\begin{tabular}{l|c|c|c|c|c}
\hline
\textbf{Method} & \textbf{L1} & \textbf{L2} & \textbf{L3} & \textbf{L4} & \textbf{Avg} \\
\hline
Basic RAG (Top-5)& 89.5 & 88.5 & 76.0 & 70.0 & 81.0 \\
\hline
Basic Agentic (Multi-turn) & 86.5 & 86.0 & 81.0 & 80.5 & 83.5 \\
\quad + Fallback Search & 90.0 & 89.5 & 84.0 & 82.5 & 87.5 \\
\quad + Chunk Delete & \textbf{91.5} & \textbf{91.0} & \textbf{88.5} & \textbf{89.0} & \textbf{90.0} \\
\hline
\textbf{Improvement over Baseline} & \textbf{+2.0} & \textbf{+2.5} & \textbf{+12.5} & \textbf{+19.0} & \textbf{+9.0} \\
\hline
\end{tabular}
}
\vspace{-5pt}
\caption{Ablation study of iterative retrieval components. Each row builds incrementally on the previous configuration.}
\label{tab:iterative_results}
\vspace{-10pt}
\end{table}

\ul{\textit{+ Fallback Search} benefits easy QAs more, i.e., L1 and L2 without retrival reasoning need, vulnerable to query drift}. Fallback search performs parallel retrieval using both the model's reformulated query and the original user query during the first retrieval turn, \ul{similar to a residual link design, significantly increases the model performance in L1 (+3.5\%) and L2 (+3.5\%), avoiding query drift}. This boost is more evident than L3 (+3.0\%) and L4 (+2.0\%).
This enhancement resulted in consistent improvements across all question levels, effectively mitigating the performance degradation observed in simple questions while maintaining benefits for complex queries.

\ul{\textit{+ Chunk Delete} instead benefits hard QA more, i.e., L3 and L4 with retrieval reasoning need, vulnerable to retrieval laziness.} As mentioned, we observed \textbf{retrieval laziness}: the agent is lazier to conduct the next retrieval when the previous round context is heavier.  
\textit{+ Chunk Delete} further boosts L3 (+4.5\%) and L4 (+6.5\%) performance, more than L1 (+1.5\%) and L2 (+1.5\%).


\begin{table}[h!]
\centering
\resizebox{1.0\columnwidth}{!}{
\begin{tabular}{l|c|c|c|c|c}
\hline
\textbf{Method} & \textbf{L1} & \textbf{L2} & \textbf{L3} & \textbf{L4} & \textbf{Avg} \\
\hline
Basic RAG (Top-5)& $89.5_{1.00}$ & $88.5_{1.00}$ & $76.0_{1.00}$ & $70.0_{1.00}$ & $81.0_{1.00}$ \\
\hline
Agentic (Qwen3)  & $\textbf{91.5}_{1.22}$& \ul{91.0}$_{1.34}$ & $\textbf{88.5}_{1.58}$ & $\textbf{89.0}_{2.22}$ & $\textbf{90.0}_{1.59}$ \\
\hline
Agentic (DeepSeek)  & $83.0_{1.86}$ & $\textbf{93.0}_{1.85}$ & $76.0_{2.02}$ & $74.0_{2.17}$ & $81.5_{1.97}$ \\
\textcolor{gray}{Agentic (DeepSeek$^*$)}  & $\textcolor{gray}{92.2}_{\textcolor{gray}{1.86}}$ & $\textcolor{gray}{96.9}_{\textcolor{gray}{1.85}}$ & $\textcolor{gray}{92.7}_{\textcolor{gray}{2.02}}$ & $\textcolor{gray}{90.2}_{\textcolor{gray}{2.17}}$ & $\textcolor{gray}{93.1}_{\textcolor{gray}{1.97}}$ \\
\hline
\end{tabular}
}
\vspace{-5pt}
\caption{Comparison with DeepSeek-R1, with accuracy as the main result and the average retrieval times executed as the subscript. The original DeepSeek results are mostly lower than Qwen3 due to dozens of questions being rejected, so we report the relative percentage of correct answers based on the questions DeepSeek can generate results for, noted as DeepSeek$^*$ in grey color.}
\label{tab:iterative_results_deepseek}
\vspace{-10pt}
\end{table}

\subsubsection{Applications on Different LLMs}

As we can observe in Table \ref{tab:iterative_results_deepseek}, no matter based on \textit{Qwen3} \cite{yang2025qwen3} or \textit{DeepSeek-R1} \cite{guo2025deepseek},  \ul{our agentic iterative strategy can generally achieve better results than basic RAG, and this may be highly due to the more frequent retrieval}: as we observe, \textit{Agentic (Qwen3)} increases retrieval by +59\%, and  \textit{Agentic (DeepSeek)} by +97\%. \ul{Moreover, in the complex questions L3 and L4, we observe twice higher frequency of retrieval.}

However, \ul{due to the high-sensitivity filtering and security protocol of \textit{DeepSeek-R1}, a lot of questions are being rejected, resulting worse performance than \textit{Qwen3}} (as shown in \textit{Agentic (DeepSeek)}); We exclude those questions and recalculate the relative accuracy rate as \textit{Agentic (DeepSeek$^*$)}, we can see that among the questions \textit{DeepSeek} is allowed to answer, the performance is very promising. Yet, this still supports our decision of deployment based on \textit{Qwen3}.


\subsection{Combine both? \re{``Cast A Big Net Multi-Times''}}
\label{sec:combine}

Overviewing the two strategies in Table \ref{tab:oneshot_results} and Table \ref{tab:iterative_results}, we can see that \ul{``different paths lead to the same goal'', both strategies, either ``cast a bigger net'' or ``cast a small net multi-times'', reach the similar accuracy of 90\% - 91\%. 
For practitioners working with government documents and sufficient token budgets, ``cast a bigger net'' is recommended due to fewer retrieval rounds and runtime; for those with limited tokens and who are comfortable with more runtime, ``cast a small net multi-times'' is recommended.}

\textbf{Combining Both}: To explore the potential synergies between One-SHOT and iterative approaches, we combined elements from both strategies. Specifically, we replaced the fixed Top-5 retriever in the Iterative Retrieval Strategy's fallback search mechanism with the Token-Constrained Top-K retriever from our One-SHOT strategy, testing various context budgets to maximize golden chunk recall in the first retrieval round.

\begin{figure}[t]
  \includegraphics[width=1\columnwidth]{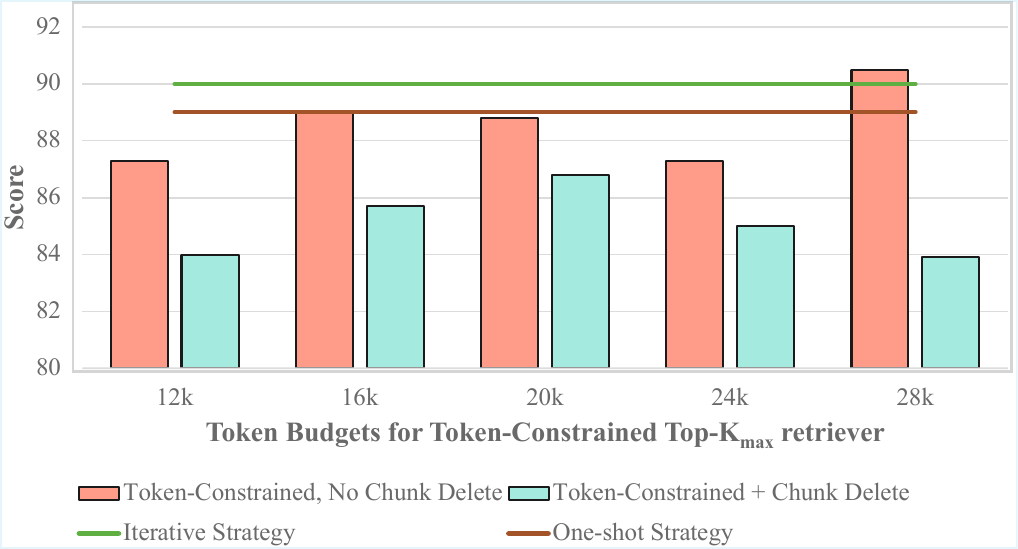}
  \vspace{-20pt}
  \caption{Performance comparison of combined strategies with different configurations.}
  \label{fig:combined_results}
  \vspace{-10pt}
\end{figure}


Surprisingly, \ul{all combined configurations underperformed the original iterative approach}. Analysis revealed that \ul{the longer working context after the first retrieval round (due to Token-Constrained Top-$K_{max}$) caused the LLM-based chunk delete tool to perform poorly, frequently removing genuinely useful chunks that were critical for answering the questions} (see Appendix~\ref{appendix:chunk_delete_failure} for a detailed case study).

When we removed the chunk delete from the combined strategy, performance improved compared to the combined strategy with chunk delete, but still fell short of the original iterative approach. It revealed that \ul{without the chunk delete, the longer working context after the first retrieval round led to severe \textbf{retrieval laziness}}. Even when information was incomplete, the model tended to avoid further retrieval operations, effectively degenerating into a One-SHOT strategy. This explains why the combined approach without chunk delete achieved scores nearly same to the pure One-SHOT.

\textbf{Analysis and Implications:} Our experiments demonstrate that the two strategies are not complementary in their current forms. The Token-Constrained Top-$K_{max}$ approach, while effective in isolation, creates longer initial contexts that interfere with the iterative retrieval process in two ways: (1) it degrades the performance of the chunk delete tool, leading to the removal of useful information, and (2) it exacerbates retrieval laziness when the chunk delete tool is removed, preventing the model from conducting necessary follow-up searches.
From cognitive science's perspective, ``People make better choices with 6 options but experience decision paralysis with 30 options'' \cite{iyengar2000choice}: this is what is happening when combining the strategies: chunk delete works fine when given 5 chunk options in the iterative strategy, but fails when being combined with Top-$K_{max}$, which usually retrieves 30+ chunks.

These findings suggest that the one-shot and iterative strategies operate under different assumptions about optimal context management and retrieval behavior. The one-shot strategy benefits from comprehensive initial retrieval within a controlled context size, while the iterative strategy relies on focused, incremental information gathering with active context management. Combining these approaches requires more sophisticated mechanisms to balance context size, relevance filtering, and retrieval motivation—an area for future research.

\section{Conclusion}
In this paper, we explored two strategies, namely “casting a bigger net” and “casting a small net multiple times”, supported by our carefully designed retrieval and reasoning modules. Both strategies demonstrated significant improvements, achieving over \textbf{+10\%} performance gains compared with basic RAG baselines when applied to complex government documents.
Our findings highlight the importance of adopting adaptive retrieval and iterative reasoning to better handle lengthy, heterogeneous government texts.
\bibliography{main}

\appendix


\section{Query Drift}
\label{appendix:query_drift}

To illustrate the query drift phenomenon discussed in Section~\ref{sec:iterative_experiments}, we provide a concrete example here:

\textbf{Original User Query:} "I have been away from my previous company for 3 years. Can I still recover my housing provident fund contributions?"

\textbf{Model-Reformulated Query:} "Regulations on housing provident fund withdrawal time limits after resignation"

\textbf{Analysis:} The original query seeks information about recovering or reclaiming housing provident fund contributions from a previous employer after a 3-year gap. However, the model's autonomous reformulation shifts the focus to withdrawal regulations and time limits, fundamentally changing the intent from recovery/reclaim to withdrawal procedures. This query drift leads to retrieval of irrelevant chunks about standard withdrawal processes rather than the specific legal provisions for recovering contributions from former employers, ultimately resulting in an incorrect or incomplete answer.

This example demonstrates how autonomous query reformulation in agentic systems can deviate from the user's original intent, particularly affecting the retrieval of relevant evidence and degrading overall system performance.

\section{Retrieval Laziness}
\label{appendix:retrieval_laziness}

To validate the retrieval laziness phenomenon discussed in Section~\ref{sec:iterative_experiments}, we conducted a controlled experiment to measure how context length affects the model's tendency to initiate follow-up retrieval calls.

\textbf{Experimental Setup:} We manually injected redundant irrelevant chunks into the initial retrieval results to reach specific context lengths while ensuring that the retrieved chunks contained only partial golden chunks—insufficient to fully answer the user's question. We then measured the probability that the model would make a subsequent \texttt{chunk\_search} call.

\begin{table}[h]
\centering
\resizebox{0.95\columnwidth}{!}{
\begin{tabular}{c|c}
\hline
\textbf{Context Length} & \textbf{Follow-up Retrieval Probability} \\
\hline
3k tokens & 95\% \\
6k tokens & 90\% \\
9k tokens & 50\% \\
12k tokens & 25\% \\
\hline
\end{tabular}
}
\caption{Retrieval laziness validation: probability of initiating follow-up retrieval calls at different context lengths.}
\label{tab:retrieval_laziness}
\end{table}

\textbf{Results:} As shown in Table~\ref{tab:retrieval_laziness}, the model's propensity to continue searching decreases dramatically as context length increases. With shorter contexts, the model correctly identifies information gaps and performs follow-up searches in the majority of cases. However, this rate drops significantly when the context becomes longer, demonstrating severe retrieval laziness in extended contexts.

This experiment confirms our hypothesis that longer working contexts exacerbate retrieval laziness, leading to premature termination of the search process even when critical information is missing.

\section{Chunk Delete Tool Failure in Combined Strategy}
\label{appendix:chunk_delete_failure}

To illustrate the chunk delete tool failure discussed in Section~\ref{sec:combine}, we provide a concrete case study demonstrating how longer working contexts impair the tool's effectiveness.

\textbf{Query:} "What was the area of flower cultivation in Kunming in the year when the first 'China·Kunming Dounan Flower Exhibition' was held?"

\textbf{Context:} The combined strategy (Token-Constrained Top-$K_{max}$ + Iterative) retrieved 10 chunks in the first round, creating a working context of 14,507 tokens. Among these chunks, one contained the crucial information that the first exhibition was held in 2023.

\textbf{Chunk Delete Tool Behavior:} When presented with this extended context, the chunk delete tool incorrectly removed 9 out of 10 chunks, including the chunk containing the correct answer about the 2023 exhibition.

\textbf{Resulting Error:} Without the correct chunk, the model provided an incorrect answer based on incomplete information.

\textbf{Analysis:} This demonstrates how extended contexts cause cognitive overload in the chunk delete tool, leading to the removal of genuinely useful information critical for accurate answers.

\end{document}